\documentclass[prd,eqsecnum,twocolumn,showpacs,preprintnumbers,amsmath,amssymb]{revtex4}

\usepackage{graphicx}

\usepackage{bm}

\newcommand{\beq}{\begin{equation}}
\newcommand{\eeq}{\end{equation}}
\newcommand{\beqs}{\begin{eqnarray}}
\newcommand{\eeqs}{\end{eqnarray}}

\begin{document}

\title{Variants of the Standard Model with Electroweak-Singlet Quarks} 

\author{Robert Shrock}

\affiliation{
C. N. Yang Institute for Theoretical Physics \\
State University of New York \\
Stony Brook, NY 11794}

\begin{abstract}

The successful description of current data provided by the Standard Model
includes fundamental fermions that are color-singlets and
electroweak-nonsinglets, but no fermions that are electroweak-singlets and
color-nonsinglets. In an effort to understand the absence of such fermions, we
construct and study {\it gedanken} models that do contain electroweak-singlet
chiral quark fields. These models exhibit several distinctive properties,
including the absence of any neutral lepton and the fact that both the $(uud)$
and $(ddu)$ nucleons are electrically charged.  We also explore how such models
could arise as low-energy limits of grand unified theories and, in this more
restrictive context, we show that they exhibit further exotic properties. 

\end{abstract}

\pacs{11.15.-q,12.10.Dm,12.60.-i}

\maketitle

\section{Introduction} 

The fundamental fermions in nature, as probed up to energies reached in
experiments so far, exhibit an intriguing asymmetry.  The asymmetry with
respect to fermion chirality is well-known.  This is evident in the fact that
the fermion content of the Standard Model (SM) is chiral with respect to its
gauge group, $G_{SM} = {\rm SU}(3)_c \times {\rm SU}(2)_L \times {\rm
U}(1)_Y$. The asymmetry that we focus on here is the fact that there are
fermions, namely the leptons, that are color-singlets but nonsinglets under the
electroweak (EW) subgroup of $G_{SM}$, $G_{EW}={\rm SU}(2)_L \times {\rm
U}(1)_Y$, but there no evidence for fermions that are singlets under $G_{EW}$
while being nonsinglets under SU(3)$_c$.  Can one understand this property of
nature at a deeper level?  To address this question, we construct and study
{\it gedanken} models that are variants of the Standard Model and that include
electroweak-singlet quarks. Our aim here is not to try to find another model
that fits current data but instead to work out properties of these {\it
gedanken} models and determine in what general ways these properties differ
from those observed in the real world.  Our methods of analysis are simply
those of quantum field theory and group theory; we do not include any results
from anthropic arguments.

   One class of variants involves the addition of a vectorlike set of
electroweak-singlet, color-nonsinglet fermions to the Standard Model.  A second
class of variants is obtained by altering the hypercharges and thus also the
electric charges of the quarks in the Standard Model so that either the
$d_R$-type or $u_R$-type quarks of each generation have $Y=Q=0$.  This can be
done in a manner consistent with constraints from anomaly cancellation so long
as one also makes corresponding changes in the charges of the leptons
\cite{nc}.  Within this class of models we discuss three particular cases. In
two of these, the electric charges of left- and right-handed Weyl components of
fermions satisfy $q_{f_L}=q_{f_R}$ and (i) $q_{d_R}=0$ or (ii) $q_{u_R}=0$.  In
the third, all fermions have $Y=0$.  We find several ways in which the
properties of such variants differ from those of the Standard Model, including
the absence of any neutral leptons and the fact that both the $(uud)$ and
$(ddu)$ nucleons are electrically charged. We then consider possible
``ultraviolet completions'' of these models \cite{uvc}. There are various
motivations, including gauge coupling unification, quark-lepton unification,
and charge quantization, to believe that the SM is a low-energy effective field
theory resulting from a grand unified theory (GUT) based on a (semi)simple
gauge group, $G_{GUT}$, with $G_{SM} \subset G_{GUT}$.  Modern grand unified
theories usually entail a supersymmetric extension of the SM \cite{susyrev},
although examples of gauge coupling unification in non-supersymmetric contexts
have also been found \cite{alt}. In the grand unified theories that we
consider, we show that such models would exhibit further exotic properties; for
example in an SU(5) theory, we find breaking of U(1)$_{em}$ by QCD quark 
condensates.

We recall certain basic properties and fix some notation.  The fermion content
of the Standard Model consists of $N_g=3$ generations of the quarks $Q^a_{n,L}
= {u^a_n \choose d^a_n}_L$, $u^a_{n,R}$, and $d^a_{n,R}$, transforming
respectively as $(3,2)_{1/3}$, $(3,1)_{4/3}$, and $(3,1)_{-2/3}$, and the
leptons $L_{n,L} = {\nu_{e_n} \choose e_n}_L$ and $e_{n,R}$ transforming as
$(1,2)_{-1}$ and $(1,1)_{-2}$.  Here $a$ is the color index, the numbers in
parentheses are the dimensions of the representations of SU(3)$_c$ and
SU(2)$_L$, the subscripts are the weak hypercharge $Y$, $n$ is the generational
index, and we use a compact notation in which $u_1 \equiv u$, $u_2 \equiv c$,
$u_3 \equiv t$, etc.  To accomodate massive neutrinos, we also include a number
$n_s$ of electroweak-singlet neutrinos, $\nu_{\ell_n,R}$ transforming as
$(1,1)_{0}$ and will usually take $n_s=N_g=3$.  We have $Q=T_3+(Y/2)$ and will
consider theories with values of $Y$ and hence $Q$ different from those in the
SM itself.  For supersymmetric extensions of the SM, we stress that our aim is
to study models with EW-singlet, color-nonsinglet matter fermions contained in
chiral superfields; of course, such models automatically include
electroweak-singlet color-adjoint fermions in vector superfields, namely the
gluinos.

\section{Model with Additional Vectorlike Fermions}

One way to construct a variant of the Standard Model with electroweak-singlet
color-nonsinglet matter fermions is simply to add a vectorlike set of
SU(2)$_L$-singlet fermions $\{f_L, \ f_R \}$, i.e. a set in which $f_L$ and
$f_R$ transform according to the same representation of SU(3)$_c$ and have the
same $Y=Q$, including some with $Y=Q=0$. If one starts with the minimal
supersymmetric Standard Model (MSSM), possibly augmented with $G_{SM}$-singlet
chiral superfields, then one would add the set of (left-handed) chiral
superfields $\{ \hat F \ , \hat F^c \}$. It is easy to see why one would not
have observed such particles at energies probed so far, since the bare fermion
mass term $m_F \bar F_L F_R + h.c.$ or corresponding superfield term $\hat F
\hat F^c$ is invariant under $G_{SM}$, and hence $m_F$ would be expected to be
of order the scale characterizing the ultraviolet completion of the theory,
such as the GUT scale. This is a special case of the general result that fields
that can form bare mass terms consistent with gauge symmetry group describing
the theory at a given scale do form such terms at this scale, and are
integrated out in the effective field theory below this scale \cite{ac}.
Rather than adding such fields to the SM or MSSM, one can, instead, change the
hypercharge assignments of the SM or MSSM fields themselves, as we discuss
next.

\section{SM with Altered Fermion Hypercharges} 

\subsection{Models with $q_{f_L}=q_{f_R}$} 

A minimal way to obtain electroweak-singlet quarks in a SM-like model or
extension thereof is to change the hypercharge assignments for the SM fermions.
To keep U(1)$_{em}$ vectorial in the simplest manner, we maintain the relations
for the electric charges
\beq
q_{f_L}=q_{f_R} \ , 
\label{qfequal}
\eeq
where $f$ runs over the quarks and leptons.  Since the $T_3=1/2$ component of
the SU(2)$_L$-doublet lepton field will have a nonzero charge for the models of
interest here, we avoid the SM notation $L_{n,L}={\nu_{e_n} \choose e_n}_L$ 
and instead write
\beq
L_L = {\ell_1 \choose \ell_2}_L \ , 
\label{ellell}
\eeq
where here and below we shall often suppress the generational index $n$.  The
altered hypercharge assignments are subject to the constraint of cancellation
of anomalies in gauged currents.  The ${\rm SU}(3)_c^3$ and ${\rm SU}(3)_c^2
{\rm U}(1)_Y$ triangle anomalies vanish because of the vectorial property of
SU(3)$_c$ and U(1)$_{em}$, The condition that the ${\rm SU}(2)_L^2 {\rm
U}(1)_Y$ triangle anomaly vanishes is
\beq
N_cY_{Q_L}+Y_{L_L}=0 \ , 
\label{su2y1anomaly}
\eeq
where we display the general $N_c$ dependence. This is equivalent to the
condition \cite{nc}
\beq
q_u = q_d+1 = \frac{1}{2}\bigg ( 1 - \frac{(2q_{\ell_2}+1)}{N_c} \bigg ) \ . 
\label{qudgeneral}
\eeq
This provides two ways to get electroweak-singlet quarks.  We discuss these for
the relevant case $N_c=3$. 

The first way entails the $Y$ assignments and corresponding SU(2)$_L$-doublets
(with electric charges in parentheses and suppressing generation indices)
\beqs
& & Y_{Q_L}=1;  \quad Q_L^a = {u^a(1) \choose d^a(0) }_L \cr\cr
& & Y_{L_L}=-3; \quad L_L   = {\ell_1(-1) \choose \ell_2(-2) }_L \ , 
\label{drssu2doublets}
\eeqs
and SU(2)$_L$ singlets having $Y_{f_R}=2q_{f_R}$, 
\beq
u_R^a(1) \ , \quad d_R^a(0) \ , \quad \ell_{1,R}(-1) \ , \quad 
                                      \ell_{2,R}(-2) \ , 
\label{drssu2singlets}
\eeq
so that $d^a_R$ is an EW singlet. We denote this case as DRS, standing for
``$d_R$ singlet''. 

The second case has SU(2)$_L$-doublets 
\beqs
& & Y_{Q_L}=-1; \quad Q_L^a = {u^a(0) \choose d^a(-1) }_L \cr\cr
& & Y_{L_L}=3;  \quad L_L   = {\ell_1(2) \choose \ell_2(1) }_L
\label{urssu2doublets}
\eeqs
and SU(2)$_L$ singlets
\beq
u_R^a(0) \ , \quad d_R^a(-1) \ , \quad \ell_{1,R}(2) \ , \quad \ell_{2,R}(1) \
, 
\label{urssu2singlets}
\eeq
so that $u^a_R$ is an electroweak-singlet (denoted case URS). 
Both the DRS and URS cases also satisfy the conditions of vanishing U(1)$_Y^3$
and $G^2 {\rm U}(1)_Y$ triangle anomalies, where $G=$ graviton.  The DRS and
URS cases correspond to cases C4$_q$ and C5$_q$ with $N_c=3$ in the
classification of Ref. \cite{nc}.

The DRS and URS models exhibit several properties that differ
from those of the Standard Model.  First, they do not have any neutral leptons.
Second, not just the proton, $p=(uud)_{J=1/2}$, but also its isospin partner
nucleon, $n=(ddu)_{J=1/2}$ (the neutron in the SM), would be charged and would
have charges $q_p$ and $q_n=q_p-1$ of the same sign:
\beq
q_p=2 \ , \quad q_n=1 \quad\quad {\rm DRS \ case}, 
\label{qpqndrs}
\eeq
\beq
q_p=-1 \ , \quad q_n=-2 \quad\quad {\rm URS \ case}.
\label{qpqnurs}
\eeq
For arbitrary $q_{\ell_2}$ it follows from eq. (\ref{qudgeneral}) that
\beq
q_p=-q_{\ell_2} \ , \quad q_n=-q_{\ell_1} \ .
\label{qperel}
\eeq
(This is true more generally for the analogues of $p$ and $n$ for higher $N_c$
\cite{nc,neutralnucleon}.)

   One can construct a supersymmetric extension of either the DRS or URS
SM-like model.  The usual Higgs mechanism in its SM or MSSM form can be
implemented for these DRS and URS models, considered in isolation.  One could
also choose one of the various scenarios for supersymmetry breaking, so that,
in the observable sector this occurs at the electroweak level, as in the MSSM.
Alternatively, as {\it gedanken} theories, one might use dynamical electroweak
symmetry breaking (EWSB) via technicolor (TC) \cite{tc} and extended
technicolor (ETC) \cite{etc}. If the residual nuclear force had the same
strength as in the real world, then the binding of nucleons to form nuclei
would be somewhat reduced because of the increased Coulomb repulsion between
nucleons resulting from the fact that both types of nucleons have nonzero
electric charges of the same sign (and indeed one of these is double the usual
proton charge in magnitude).  This would tend to destabilize some nuclei that
are stable in the real world. Although both members of the nucleon isodoublet
are charged, there are spin-1/2 baryons that are neutral in this model.  For
the DRS and URS cases these include, for example, the spin-1/2 baryons
\beq
(dds), \quad (dss), \quad (ddb), \quad etc. \quad  ({\rm DRS \ case}) 
\label{dds}
\eeq
\beq
(uuc), \quad (ucc), \quad (uut), \quad etc. \quad  ({\rm URS \ case}) \ . 
\label{uuc}
\eeq
If one keeps the masses of the quarks equal or similar to their values in the
real world, then these are heavier than the nucleons and would beta decay.  We
shall show below how the matter fermion content of the DRS and URS models can
arise from a grand unified theory, where their structure is more tightly
constrained.

\subsection{Model with all Fermions Having $Y=0$}

A different modification of the SM with electroweak-singlet quarks that is
allowed by anomaly constraints is for each fermion generation to have the
SU(2)$_L$ doublets (again suppressing the $n$ index) 
\beqs
& & Y_{Q_L}=0; \ Q_L^a = {u^a(1/2) \choose d^a(-1/2)}_L \ , \cr\cr
& & Y_{L_L}=0; \ L_L = {\ell_1(1/2) \choose \ell_2(-1/2)}_L 
\label{yleftz}
\eeqs
and SU(2)$_L$ singlets $\{f_R\}$ with 
\beq
Y_{f_R}=q_{f_R}=0 \quad \forall \ f_R \ . 
\label{yrightz}
\eeq
To keep SU(3)$_c$ vectorial, the set $\{f_R\}$ includes two color triplets for
each generation, which we denote $\eta^a_R$ and $\eta'^a_R$.  (We avoid
denoting these as $u^a_R$ and $d_R$ since they have different electric charges
than $u^a_L$ or $d^a_L$.)  The remainder of the set $\{f_R\}$ is comprised of
(two or some other number of) $G_{SM}$-singlets.  We denote this model with the
matter fermion content in Eqs. (\ref{yleftz}) and (\ref{yrightz}) as the YZ
($Y$ zero) case; it corresponds to the case C2$_{q,sym}$ (equivalently
C2$_{\ell,sym}$) in the classification of Ref. \cite{nc}.  Although SU(3)$_c$
(with other interactions turned off) and U(1)$_{em}$ (with other interactions
turned off) are vectorial symmetries in the YZ model, this occurs in a
``twisted'' manner, in which there is not a 1-1 correspondence between a
left-handed Weyl field and a right-handed Weyl field with the same color and
charge \cite{unt}.  From Eq. (\ref{qperel}), it follows that both members of
the nucleon isodoublet are charged:
\beq
q_p = -q_n = \frac{1}{2}  \quad\quad ({\rm YZ \ case}) \ . 
\label{qpqnyz}
\eeq

If one considers the YZ model in isolation without trying to construct an
ultraviolet completion, then one can include a $I=1/2$, $Y=1$ SM Higgs field
$\phi$ or, in an MSSM context, $I=1/2$ $Y=\pm 1$ Higgs chiral superfields.
With either of these one can break ${\rm SU}(2)_L \times {\rm U}(1)_Y$ to
U(1)$_{em}$ via Higgs vacuum expectation values (VEV's). However, one cannot
construct $G_{SM}$-invariant Yukawa couplings and use these to generate masses
for the matter fermions.  For example, the Yukawa term $\bar Q_{a,n,L}
f^a_{n',R} \phi + h.c.$ is forbidden by U(1)$_Y$ gauge invariance, since it
transforms as a $Y=1$ operator.  Assigning any value of $Y$ other than $\pm 1$
to the Higgs field(s) would not allow EWSB, since the Higgs would not have any
neutral components.  

In this YZ model, QCD confines and spontaneously breaks chiral symmetry. 
The most attractive channel for condensate formation, $3 \times \bar 3 \to
1$, yields the condensates (suppressing $n$ indices) $\langle \bar u_{a,L} \, f
^a_R\rangle$ and $\langle \bar d_{a,L} \, f^a_R\rangle$, where $f_R$ refers to
$\eta_R$ or $\eta'_R$. Without loss of generality, we can write these as
$\langle \bar u_{a,L} \, \eta^a_R\rangle$ and $\langle \bar d_{a,L} \,
\eta'^a_R\rangle$.  Since $q_{u_L}=1/2$, $q_{d_L}=-1/2$, and
$q_{\eta_R}=q_{\eta'_R}=0$, these condensates break not just SU(2)$_L$, but
also U(1)$_{em}$.  This model is thus strikingly different from the real world.
We shall show below how the matter fermion content of the YZ model (but not 
$I=1/2$, $Y=\pm 1$ Higgs field(s)) arises naturally as a low-energy effective
field theory if one requires electroweak-singlet fermions in an SU(5) GUT.

\section{Grand Unification in SU(5)}

We now analyze electroweak-singlet quarks in the context of grand unified
theories.  Much modern work on grand unified theories has focused on meeting
constraints from proton decay and deriving models from a presumed underlying
string theory. Our purpose here is somewhat different; we are not trying to
account in detail for the experimentally observed values of gauge couplings or
limits on proton decay. Instead, we wish to explore the properties of {\it
gedanken} grand unified theories containing electroweak-singlet quarks,
accepting that these would entail changes in the measured values of
$\sin^2\theta_W$, etc.  We first consider the case where the GUT group has the
minimal rank, namely 4, the same as $G_{SM}$. For this case, the canonical
choice is $G_{GUT}={\rm SU}(5)$ \cite{gg}.  One assigns the left-handed matter
fermions of each generation to a $\bar 5$ and $10$ representation.  Under ${\rm
SU}(3)_c \times {\rm SU}(2)_L$ these decompose as $\bar 5=(\bar 3,1) \oplus
(1,2)$ and $10 = (\bar 3,1) \oplus (3,2) \oplus (1,1)$.  In order to make the
$(\bar 3,1)$ in the $\bar 5$ of SU(5) an EW-singlet (anti)quark, we assign it
zero hypercharge. In terms of the the equivalent $5_R$, we write
\beq
\psi_R = \left ( \begin{array}{c} 
               \eta^a \\
               L^c \end{array} \right )_R
\label{5right}
\eeq
where $a$ is again the color index and $Y_{\eta_R}=0$. As before, we use the
symbol $\eta_R$ rather than $u_R$ or $d_R$ for this quark because it will not
have the charge of either $u_L$ or $d_L$.  If the GUT group is SU(5), then $Y$
(and hence $Q$) are (linear combinations of) generators of the Lie algebra of
SU(5) and hence satisfy ${\rm Tr}(Y)={\rm Tr}(Q)=0$.  Therefore, $Y_{L^c_R}=0$,
and $L_L={\ell_1(1/2) \choose \ell_2(-1/2)}_L$ (charges listed in parentheses).
Consequently, as operators, $Y = {\rm diag}(0,0,0,0,0)$ and thus
\beq
Q = {\rm diag}(0,0,0,1/2,-1/2) \ . 
\label{qyz} 
\eeq
These operators are to be contrasted with the forms in conventional SU(5)
\cite{gg}, 
\beq
Y_{conv.}  = {\rm diag}(-2/3,-2/3,-2/3,1,1) 
\label{yconv}
\eeq
and thus
\beq
Q_{conv.} = {\rm diag}(-1/3,-1/3,-1/3,1,0) \ . 
\label{qconv}
\eeq
It follows that the $Y$ assignments in the 10 of SU(5) in the present case are
\beq
10 = (\bar 3,1)_0 \oplus (3,2)_0 \oplus (1,1)_0 \ , 
\label{10yz}
\eeq
with respective component fields
\beq
\eta'^c_{a,L}(0) \ , \quad {u^a(1/2) \choose d^a(-1/2)}_L \ , \quad \chi^c_L(0)
\ . 
\label{10fields}
\eeq
Thus, for each generation (suppressing the generation index) the fermions from
the $\bar 5_L$ and $10_L$ comprise the representations
\beqs
& & Q^a_L = {u^a(1/2) \choose d^a(-1/2)}_L: \ \ (3,2)_0 \ , \cr\cr
& &   L_L = {\ell_1(1/2) \choose \ell_2(-1/2)}_L: \ \ (1,2)_0 
\label{su2doublets}
\eeqs
and
\beq
\eta^a_R(0), \ \eta'^a_R(0): \ \ (3,1)_0 \ , \quad \chi_R(0): \ \ (1,1)_0 \ . 
\label{su2singlets}
\eeq
This therefore yields a YZ-type model.  There could also be SU(5)-singlet
matter fermions.

One envisions that the SU(5) gauge symmetry is spontaneously broken to $G_{SM}$
at the GUT scale, $M_{GUT}$.  The resultant theory below $M_{GUT}$ has several
properties that are quite different from those of the real world. First, since
all of the particles have $Y=0$, the effective gauge group is just ${\rm
SU}(3)_c \times {\rm SU}(2)_L$, without a U(1)$_Y$ factor.  Thus, in this model
$Q=T_3$ and U(1)$_{em}$ is a subgroup of SU(2)$_L$.  Second, as noted above,
although the SU(3)$_c$ and U(1)$_{em}$ gauge interactions are individually
vectorial, this is realized in a manner different from that of the SM. Third,
one cannot construct a SM- or MSSM-type Higgs sector in this theory because the
operator $Q$ has no color-singlet, SU(2)$_L$-doublet, electrically neutral
component.  For example, the Higgs SU(2)$_L$ doublet contained in an SU(5) 5 of
Higgs is $\phi={\phi_1(1/2) \choose \phi_2(-1/2)}$.  A VEV for either component
of this Higgs doublet breaks SU(2)$_L$ completely, including its U(1)$_{em}$
subgroup.  For the same reason, usual SM-type Yukawa couplings and their
supersymmetric extensions are not possible in this theory.  The
(SU(3)$_c$-invariant) mass terms that one might consider for the quarks,
\beq
\sum_{n,n'} \bar q_{a,n,L} M^{(q)}_{nn'} f^a_{n',R} + h.c.
\label{qmass}
\eeq
with $q=u,d$ and $f=\eta, \ \eta'$, break SU(2)$_L$ and hence also U(1)$_{em}$.
The same is true of the lepton Dirac mass terms
\beq
\sum_{n,n'} \bar \ell_{j,n,L} M^{(\ell_j)}_{nn'} \chi^a_{n',R} + h.c.\ , \quad 
\label{ellmass}
\eeq
where $j=1,2$. The SU(2)$_L$-doublet leptons can have the $G_{SM}$-invariant
Majorana bare mass terms
\beq
\sum_{n,n'} \epsilon_{ij} L_{n,L}^{i \ T} C M^{(L)}_{nn'}L^j_{n',L} + h.c. 
\label{mell}
\eeq
where $i,j$ denote SU(2)$_L$ indices.  The structure of this operator implies
that $M^{(L)}_{nn'}=-M^{(L)}_{n'n}$, so for the relevant case of odd $N_g$
there is at least one zero eigenvalue, i.e., a massless charged lepton at this
level.  Via diagrams involving the exchange of GUT-scale gauge bosons, the
proton and $\ell_2^c$ will mix, as will the $(ddu)_{J=1/2}$ nucleon and
$\ell_1^c$, which will give extremely small masses to these leptons.  The
effect of the very low-mass charged unconfined leptons is reduced by the fact
that U(1)$_{em}$ is broken, as we discuss next.  The $\chi_{n,R}$'s could have
bare Majorana mass terms $\sum_{n,n'}\chi_{n,R}^T C M^{(\chi)}\chi_{n',R} +
h.c.$.

Since the SU(3)$_c$ gauge interaction is asymptotically free, as the energy
scale $\mu$ decreases below $M_{GUT}$, the SU(3)$_c$ coupling grows.  As $\mu$
decreases through $\Lambda_{QCD}$ and $\alpha_s = g_s^2/(4\pi)$ reaches values
of order unity, the QCD sector exhibits confinement. Since mass terms for the
quarks would violate U(1)$_{em}$, we take them to be massless.  Then in the
limit where one neglects electroweak interactions, the theory has a global
flavor symmetry
\beq
G_{fl} = {\rm SU}(2N_g)_L \times {\rm SU}(2N_g)_R 
\label{ggl}
\eeq
The QCD interaction spontaneously breaks $G_{fl}$ by the formation of the
bilinear quark condensates (in the $3 \times \bar 3 \to 1$ channel)
\beq
\sum_{n,n'} \langle \bar q_{a,n,L} f^a_{n',R}\rangle + h.c., 
\label{qcond}
\eeq
where $q=u,d$ and $f=\eta, \ \eta'$.  As with the mass terms, these condensates
are invariant under SU(3)$_c$ but break SU(2)$_L$ and hence U(1)$_{em}$. This
is the same as the YZ model, now seen in a GUT context.

\section{Grand Unification in SO(10)}

One can construct grand unified theories with electroweak-singlet quarks that
avoid the breaking of U(1)$_{em}$ by using $G_{GUT}={\rm SO}(10)$ \cite{so10}
and taking advantage of the additional freedom of having $Y$, and hence $Q$, be
generators of SO(10) but not of SU(5).  SO(10) models in which $Y$ and $Q$ are
not generators of SU(5) were constructed in Ref. \cite{xy} with conventional
quark and lepton charges and the $u^a_R$ assigned to the $5_R$, so that $Y={\rm
diag}(4/3,4/3,4/3,1,1)$.  We will use this freedom in a different way here.  We
denote $\tilde Y$ as the generator of SU(5) that commutes with SU(3)$_c$ and
SU(2)$_L$. Now SO(10) contains, as a maximal subgroup, ${\rm SU}(5) \times {\rm
U}(1)_X$, and the spinor 16 of SO(10) transforms as $16 = 1_5 \oplus \bar
5_{-3} \oplus 10_1$, where subscripts denote $X$ values.  We set
\beq
\tilde Y = aY + bX \ . 
\label{yeq}
\eeq
Since the electroweak-singlet quark is assigned to the first $N_c=3$ components
of $\psi_R$, it follows that for the 5 of SU(5),
\beq
Y={\rm diag}(0,0,0,y,y) \ ,  \quad y=Y_{L^c_R}=-Y_{L_L} \ . 
\label{ygen}
\eeq
Hence ${\rm Tr}(\tilde Y)=2ay+15b$. We can take $a=1$, and solve to get
$b=-2y/15$, so that
\beq
\tilde Y = Y - \frac{2y}{15}X \ .
\label{ytildegen}
\eeq

For the DRS case, with $d_R$ in $\psi_R$ and $Y_{L_L}=-3$, we thus have $\tilde
Y=Y-(2/5)X$. For the 5 of SU(5) this is
\beq
DRS: \quad \tilde Y = \frac{1}{5}{\rm diag}(-6,-6,-6,9,9) \ . 
\label{ytildedrs}
\eeq
For a representation $R$ with a given value of $X$, one then calculates $Y$ and
$Q$ by using Eq. (\ref{ytildegen}).  For the 5 of SU(5), $Y={\rm
diag}(0,0,0,3,3)$ and $Q={\rm diag}(0,0,0,2,1)$.  The components of the 10 of
SU(5) have the $Y$ values indicated,
\beq
DRS: \quad  10_L: \ (\bar 3,1)_{-2} \oplus (3,2)_1 \oplus (1,1)_4 \ , 
\label{10drs}
\eeq
with component fields given by $u^c_{a,L}(-1)$, ${u^a(1) \choose d^a(0)}_L$
and $(\ell_2^c)_L(2)$.  The remaining component of the 16 of SO(10) is an SU(5)
singlet, $(\ell_1^c)_L(1)$.  These fields thus comprise the set in
Eqs. (\ref{drssu2doublets})-(\ref{drssu2singlets}).

For the URS case, with $u_R$ in $\psi_R$ and $Y_{L_L}=3$, we have 
$\tilde Y=Y+(2/5)X$. For the 5 of SU(5) this is
\beq 
URS: \quad \tilde Y = \frac{1}{5}{\rm diag}(6,6,6,-9,-9) \ . 
\label{ytildeurs}
\eeq
Thus for the 5 of SU(5), $Y={\rm diag}(0,0,0,-3,-3)$ and
$Q={\rm diag}(0,0,0,-1,-2)$.  The components of the 10 of SU(5) have the $Y$
values indicated, 
\beq
URS: \quad 10_L: \ (\bar 3,1)_{2} \oplus (3,2)_{-1} \oplus (1,1)_{-4}
\label{10urs}
\eeq
with component fields given by $d^c_{a,L}(1)$, ${u^a(0) \choose d^a(-1)}_L$ and
$(\ell_1^c)_L(-2)$.  The remaining component of the 16 of SO(10) is an SU(5)
singlet, $(\ell_2^c)_L(-1)$. These fields make up the set in Eqs.
(\ref{urssu2doublets})-(\ref{urssu2singlets}).  

Using the relation $\sin^2 \theta_W = {\rm Tr}_R(T_3^2)/{\rm Tr}_R(Q^2)$, 
we find
\beq
\sin^2 \theta_W = \frac{1}{2(1+3Y_{Q_L}^2)}  \quad {\rm at \ } M_{GUT}. 
\label{sinsq}
\eeq
Thus, at $M_{GUT}$, $\sin^2\theta_W \le 1/2$, and this value is reached for the
$Y$ choices leading to the YZ low-energy field theory.  The $Y$ choices leading
to the DRS or URS models yield $\sin^2 \theta_W = 1/8$ at $M_{GUT}$.

In these SO(10)-based DRS and URS models the 16-dimensional spinor
representation has no color-singlet, SU(2)$_L$-doublet, electrically neutral
entries, in contrast to both conventional \cite{so10} and `flipped'' \cite{xy}
SO(10) models, and the same follows for the $\bar 5$ and 10 representations of
SU(5) arising from this spinor.  A Higgs (super)field that could give rise to
electroweak symmetry breaking is thus problematic.  One could consider adding a
(super)field transforming as a vector representation of SO(10).  Under ${\rm
SU}(5) \times {\rm U}(1)_X$ the 10 of SO(10) decomposes as $10 = 5_{-2} \oplus
\bar 5_2$. In the DRS and URS cases the 5 of Higgs resulting from this would
have charges $(-1,-1,-1,1,0)$ and $(1,1,1,0,-1)$, respectively. Although these
Higgs fields thus contain color-singlet, SU(2)$_L$-doublet, neutral entries,
they can form bare, SO(10)-invariant mass terms with masses naturally of order
the GUT scale and hence would be integrated out of the low-energy theory
operative below this scale.  Moreover, if such mass terms were not present,
then color-triplet Higgs components in the Higgs 10 of SO(10) could contribute
to overly rapid nucleon decay.  For the purposes of our further analysis, we
assume that GUT-scale mass terms are present for Higgs (super)fields
transforming as the 10 of SO(10).

  If one were to depart from this simple GUT and adjoin to the theory a
technicolor sector, then one could use this to break electroweak symmetry at
the 250 GeV scale (where in this {\it gedanken} model we would accept resultant
technicolor modifications of $Z$ and $W$ propagators) \cite{etcgut}.  However,
if we consider the theory by itself, without such an addition, then we may ask
if this resultant theory would break electroweak symmetry.  The answer is yes,
and this breaking is dynamical.  There would not be Higgs-based Yukawa
couplings or TC/ETC contributions to give fermions masses. In the absence of
these and in the limit where one turns off electroweak interactions, the QCD
sector would have the global chiral symmetry $G_{fl}$ in Eq. (\ref{ggl}). (Note
that there would not be any strong CP problem in QCD because the massless
quarks would allow one to rotate away the $\bar\theta$ angle.) Because the
SU(3)$_c$ gauge interaction is asymptotically free, $\alpha_s$ increases as the
energy scale decreases below $M_{GUT}$, reaching O(1) at $\Lambda_{QCD}$, at
which scale this interaction produces the bilinear quark condensates in the $3
\times \bar 3 \to 1$ channel. We now turn the electroweak interactions back on.
By vacuum alignment arguments, the condensates preserve U(1)$_{em}$ and the
diagonal, vectorial subgroup ${\rm SU}(2N_g)_V$ of $G_{fl}$, thereby producing
$(2N_g)^2-1$ Nambu-Goldstone bosons (NGB's). Without loss of generality, we can
define the ordering of the generational bases for left- and right-handed quarks
so that these condensates take the form (summed on $a$, not on $n$),
\beqs
& & \langle \bar u_{a,n,L} u^a_{n,R}\rangle + h.c. \cr\cr
& & \langle \bar d_{a,n,L} d^a_{n,R}\rangle + h.c., \quad 1 \le n \le N_g \ . 
\label{condensates}
\eeqs
These condensates transform with weak $I=1/2$, $|Y|=1$ and break ${\rm SU}(2)_L
\times {\rm U}(1)_Y$ to U(1)$_{em}$, so three of the NGB's are absorbed to
produce longitudinal components and masses for the $W^\pm$ and $Z$, leaving the
remaining $4(N_g^2-1)$ (P)NGB's in the spectrum.  Denoting $f_\pi$ as the
generalized pion decay constant and $g$ and $g'$ as the SU(2)$_L$ and U(1)$_Y$
gauge couplings, one has $m_W^2 = N_g g^2 f_\pi^2 N_g/4$ and $m_Z^2 = N_g
(g^2+g'^2)f_\pi^2/4$, satisfying $m_W^2 = m_Z^2 \cos^2\theta_W$.  The
realization that QCD quark condensates break electroweak symmetry and that
three of the resultant NGB's would be absorbed to give the $W^\pm$ and $Z$
masses was, indeed, one of the main motivations for the original development of
technicolor \cite{tc}. 

   The confinement and spontaneous chiral symmetry breaking in the QCD sector
generates dynamical, constituent masses of order $\Lambda_{QCD}$ for the
quarks. Provided that the quarks have zero hard masses, the resultant
constituent masses would be equal for $u$-type and $d$-type quarks, up to
electromagnetic corrections.  We assume that if one begins with a
supersymmetric grand unified theory, the supersymmetry is broken at a higher
scale.  The spectrum of the theory depends sensitively on the number of matter
fermion generations.  In the hypothetical case $N_g=1$, since the NGB's (pions)
are absorbed by the $W^\pm$ and $Z$, the low-lying hadron spectrum would be
comprised of the isovector $\rho$ and isoscalar $\omega$, the nucleons, the
(non-NGB) isoscalar pseudoscalar analogue of $\eta'$, and so forth.  For
$N_g=2$ or $N_g=3$, the spectrum would be qualitatively different because of
the (i) residual (P)NGB's and (ii) electrically neutral, spin-1/2, ground state
baryons.  Indeed, the absence of hard, current-quark masses would mean that, up
to electromagnetic effects, the various ground-state baryons of a given spin
would be essentially degenerate. Thus, although the $(uud)$ and $(ddu)$ baryons
would be charged, there would be the spin-1/2 ground-state baryons listed in
Eqs. (\ref{dds}) and (\ref{uuc}) for the respective DRS and URS cases.

Concerning the charged baryons, one expects a Coulombic energy contribution
roughly proportional to $q^2/R$ to a hadron of charge $q$ and size $R$. Hence,
in the absence of hard quark masses, for $N_g=1$, the $p$ and $n$ would be the
lightest baryons, while for $N_g =2,3$ the lightest baryons would be the
neutral ones in the respective Eqs. (\ref{dds}) and (\ref{uuc}).  Among
ground-state baryons, electromagnetic mass differences would be of order
$\simeq \alpha_{em} \Lambda_{QCD}$, i.e., a few MeV if one continues to
take $\Lambda_{QCD}$ to have its real-world value of $\sim 200$ MeV.  Since the
size of the proton $p$ and its isospin partner nucleon $n$ would be essentially
the same, and since $q_p = 2$ and $q_n=1$ for the DRS case, one infers that
\beq
m_p > m_n  \quad\quad {\rm DRS \ case} \ . 
\label{mpmndrs}
\eeq
For the URS case, since $q_p=-1$ while $q_n=-2$, one has
\beq
m_n > m_p  \quad\quad {\rm URS \ case} \ . 
\label{mpmnurs}
\eeq
Given that the leptons have very
small masses (see below), the following beta decays would occur:
\beq
p \to n \ell_1 \ell_2^c  \quad\quad ({\rm DRS}) 
\label{pbetadrs}
\eeq
and 
\beq
n \to p \ell_2 \ell_1^c  \quad\quad ({\rm URS}) \ . 
\label{nbetadrs}
\eeq

In each of these cases, one could make the formal observation that the lighter
nucleon could form a neutral Coulombic bound state that are stable with respect
to strong and weak decays, namely
\beq
[n(1)\ell_1(-1)] \quad\quad ({\rm DRS})
\label{hdrs}
\eeq
and
\beq
[p(-1)\ell_2(1)] \quad\quad ({\rm URS}) \ , 
\label{hurs}
\eeq
where we have indicated charges in parentheses and implicitly refer to the
lightest mass eigenstates in the relevant interaction eigenstates $\ell_j$.
Also, formally, there could be stable three-body leptonic Coulomb bound states,
\beqs
& & [\ell_2(-2) \ell_1^c(1)\ell_1^c(1)]  \quad\quad ({\rm DRS}) \ , \cr\cr
& & [\ell_1(2) \ell_2^c(-1)\ell_2^c(-1)]  \quad\quad ({\rm URS}) \ . 
\label{leptonicstate}
\eeqs

However, in the absence of Lagrangian mass terms for the matter fermions, both
the DRS and URS models resulting from this SO(10) GUT have charged, unconfined
$\ell_1$ and $\ell_2$ leptons with zero Lagrangian masses. There is mixing
between the $p$ and $\ell_2^c$ and, separately, mixing between the $n$ and
$\ell_1^c$. These mixings generate nonzero, although extremely small, masses
for these leptons.  To illustrate this, let us take the DRS case for
definiteness (similar statements hold for the URS case) and $N_g=1$ for
simplicity.  Here the GUT gauge boson sector includes bosons
$(Y^a(-1),X^a(-2))$ transforming as $(3,2)_{-3}$ and their adjoints.  A
tree-level amplitude contributing to proton decay is $u+u \to d^c+\ell_2^c$,
mediated by the $s$-channel exchange of a $X^\dagger_a$.  The corresponding
$uud \ell_2$ operator also gives rise to the mixing $u+u+d \leftrightarrow
\ell_2^c$. The matrix element of this operator between the states $|p\rangle$
and $|\ell_2^c \rangle$ is $\delta \equiv Amp(p \leftrightarrow \ell_2^c)
\simeq (g_{GUT}^2/M_{GUT}^2)\Lambda_{QCD}^3$, where $g_{GUT}$ is the GUT gauge
coupling. Diagonalizing the corresponding $2 \times 2$ mixing matrix, one finds
a negligibly small shift in the proton mass $m_p$ and a nonzero mass
\beq
m_{\ell_2} = \frac{|\delta|^2}{m_p} \ .  
\label{me}
\eeq
Using $\Lambda_{QCD}=200$ MeV and the illustrative values $\alpha_{GUT} \simeq
1/24$ and $M_{GUT} \simeq 10^{16}$ GeV, we have $m_{\ell_2} \simeq 10^{-72}$
GeV.  A similar mass is generated for $\ell_1$.  Since these masses are so
small, they are obviously quite sensitive to additional ingredients in an
ultraviolet completion of the theory and should only be considered as
illustrative.  Formally, an isolated Coulombic state such those in
Eqs. (\ref{hdrs}) or (\ref{hurs}) would have a size set by the Bohr radius $a
\sim (\alpha_{em}m_{\ell_j})^{-1}$, $j=1,2$.  However, this is only formal,
since with the tiny $\ell_j$ masses, this size would be many orders of
magnitude larger than the present size of the (real-world) universe.
Unconfined charged leptons with such small masses would cause an infrared
instability in the theory, so that the above-mentioned Coulombic bound states
would be replaced by a plasma, similar to the situation in the hypothetical
case of SM with conventional fermion charges but without a Higgs field
\cite{q}.  

In view of these various results, we may conclude that even though our SO(10)
GUT constructions yielding DRS and URS models as low-energy effective field
theories avoid the breaking of U(1)$_{em}$ that afflicted the YZ model, they
still exhibit exotic properties and striking differences as compared with
the real world.  Some further remarks may be in order. In the absence of a
usual SM Higgs or the pair of MSSM Higgs chiral superfields, the perturbatively
calculated partial wave amplitudes for the scattering of longitudinally
polarized vector bosons in the DRS and URS models would exceed unitarity upper
bounds at a center-of-mass energy somewhat above the EWSB scale.  The
unitarization of these amplitudes would depend on the ultraviolet completion of
the theory.  If the only source of EWSB is QCD, then this unitarization would
involve exchanges of the scalar and vector hadrons of QCD at the scale of O(1)
GeV.  If one were to adjoin a TC/ETC sector to obtain EWSB at the usual
physical scale of 250 GeV, then this unitarization would involve exchanges of
technihadrons.  One could also consider larger grand unified groups.  In
particular, we have studied electroweak-singlet quarks in the context of a GUT
based on the gauge group E$_6$, which contains SO(10) as a subgroup, and have
obtained similar conclusions.

Our finding that the DRS and URS models are more tightly constrained when
considered as low-energy effective field theories resulting from an SO(10)
grand unified theory than when considered in isolation is understandable, since
the larger structure of a GUT provides a more predictive theoretical framework.
This is analogous to the fact that the ratios of gauge couplings for the three
factor groups in the Standard Model are essentially arbitrary when this theory
is considered in isolation, but are predicted when it (or its supersymmetric
extension) is embedded in a grand unified theory.  In principle, one might
further extend the present analysis of electroweak-singlet quarks to the case
of GUT's containing an extension of the SM with $N_c$ different from 3.
However, although one can satisfy anomaly constraints in an $N_c$-extended
Standard Model, the natural embedding of such a theory in an ${\rm
SO}(2(N_c+2))$ grand unified theory would require 
\beq
2^{N_c+1}=4(N_c+1) \ , 
\label{ncp}
\eeq
which only has a solution for $N_c=3$ \cite{nc,sim}.  Accordingly, we do not
pursue such an $N_c$-generalization here.

\section{Conclusions} 

In this paper we have sought to gain a deeper understanding of one property of
the Standard Model, namely the absence of electroweak-singlet quarks. For this
purpose we have constructed and studied {\it gedanken} models that are similar
to the Standard Model but do contain electroweak-singlet quarks.  We have found
that these models exhibit properties fundamentally different from those of the
Standard Model, including the absence of neutral leptons and the fact that both
the $(uud)$ and $(ddu)$ nucleons are charged.  Furthermore, working in the
context of a grand unified theory, we have shown that (i) an SU(5) theory with
electroweak-singlet quarks leads to a low-energy field theory which, among
other things, violates U(1)$_{em}$; and (ii) SO(10) grand unified theories in
which $Y$ and $Q$ are generators of SO(10) but not SU(5) can avoid breaking of
U(1)$_{em}$, but still generically lead to low-energy effective field theories
quite different from the Standard Model.

This research was partially supported by the grant NSF-PHY-06-53342.


\begin{thebibliography}{99}
\bibitem{uvc}
Although we use the common term ``ultraviolet completion'', a preferable term
might be ``ultraviolet extension'', since such theories could well have 
further new structure at yet higher energies.
\bibitem{neutralnucleon}
%
One may ask, for general $N_c$, which choices of matter fermion charges could
yield a neutral nucleon.  In addition to the conventional SM set of charges,
there is one other SM-like set: $Y_{Q_L}=-1/N_c$, $Y_{L_L}=1$, and
$q_{f_L}=q_{f_R}$, resulting in $q_p=0$, $q_n=-1$ (case C4$_\ell$ in
\cite{nc}).  For $N_c=3$ this set can be embedded in an SU(5) GUT with $5_R$
containing $u^a_R(1/3)$ and $L^c_R$, where $L_L={\ell_1(1) \choose
\ell_2(0)}_L$, $10_L$ containing $Q^a_L={u^a(1/3) \choose d^a(-2/3)}_L$,
$d^c_{a,L}(2/3)$, and $\ell_{1,L}^c(-1)$, with $Y={\rm
diag}(2/3,2/3,2/3,-1,-1)$ and a possible SU(5)-singlet $\ell^c_{2,L}(0)$.
\bibitem{unt}
A model with $Y_{Q_L}=Y_{L_L}=0$ and conventionally vectorial SU(3)$_c$ and 
U(1)$_{em}$ would have SU(2)$_L$-singlet fields $u^a_R(1/2), \ d^a_R(-1/2), 
\ \ell_{1,R}(1/2), \ \ell_{2,R}(-1/2)$ rather than Eq. (\ref{yrightz}).
\end{thebibliography}
\end{document}